# Pulse width modulation technique with harmonic injection in the modulating wave and discontinuous frequency modulation for the carrier wave to reduce vibrations in asynchronous machines

Antonio Ruiz-Gonzalez[1], Mario Meco-Gutierrez[1], Juan-Ramon Heredia-Larrubia[2], Francisco Perez-Hidalgo[1], Francisco Vargas-Merino[1]

[1]University of Malaga EII, Electrical Engineering Dept. Doctor Ortiz Ramos s/n 29071 Málaga (SPAIN)

[2]University of Malaga EII, Electronic Technology Dept. Doctor Ortiz Ramos s/n 29071 Málaga (SPAIN)

**Abstract:** A new carrier-based pulse-width modulation (PWM) technique to control power inverters is presented in this paper. To generate the output waveform, this technique compares a harmonic-injection modulating wave and a frequency-modulated triangular carrier wave. The instantaneous frequency for the carrier wave is adjusted according to a periodic function synchronized with the fundamental term of the modulating wave. The main motivation for using this technique compared to a classic PWM sinusoidal technique revolves around the reduction of total harmonic distortion, the reduction of the distortion factor and the shift of temporal harmonics to higher frequencies for any modulation frequency order. Experimental results show that it is possible to optimize the time harmonics generated to minimize vibrations produced by an induction motor when it is fed with a DC/AC converter controlled by the proposed control strategy. This is made possible by using a control parameter that modifies the instantaneous frequency of the carrier wave without modifying the number of pulses per period of the modulating wave, i. e. the mean value of the carrier wave frequency. The proposed technique is applied to an open loop-controlled inverter that operates an induction motor, helping to reduce the vibration levels produced.

## Nomenclature

- $A_M$    control parameter to adjust the maximum frequency of the carrier wave
- $K$    truncated level of the function that controls the carrier wave frequency
- $\omega_i$    instantaneous pulsation of carrier wave
- $\omega_m$    pulsation of the modulating wave
- $M$    frequency modulation order
- $\overline{M}$    average value of the frequency modulation order

## 1. Introduction

The traditional carrier-based pulse-width modulation (PWM) technique is the most common method of controlling the pulses width which are applied to the electronic switches of the DC/AC converter and it is adopted for both multilevel [1-2] and 3-levels inverters [3]. Although this technique is easy to implement, its main disadvantage is that it generates a fundamental term with a small amplitude and a significant number of undesired harmonics for the 3-levels inverters. These harmonics increase losses, vibrations, and noises, thereby degrading the performance of the system.

In technical literature, different methods to reduce harmonics can be found. Using the space-vector–based hybrid PWM technique, it is possible to reduce the current ripple [4-5]. Another method is to use a hybrid technique combining both the generated-carrier–based PWM and the space-vector–based hybrid PWM [6]. Researchers have also developed multilevel techniques to reduce the total harmonic distortion (THD) and distortion factor (DF) that could be applied to reduce the thermal stress, acoustic noise and vibration levels produced by avoiding excitation of resonance frequencies, or space harmonics for sensitive frequencies (with high winding factors) in the lowest mechanical vibratory modes of the induction motor fed [7-14]; and to reduce common-mode voltage that occurs in five-phase converters [15]. However, these methods present the disadvantage of greater complexity of the inverter structure.

Another solution is to use pulse-frequency modulation (PFM). This strategy is based on the frequency modulation of the triangular carrier using a periodic signal [16]-[17]. A good approach had been to use a trapezoidal modulating wave when this one is compared with a carrier (triangular or sinusoidal) during the time interval that the slope of the trapezoidal wave is not null [18]-[19]. This approach acts as a slope-modulation strategy for the generated PWM.

Other works in the reduction line of electric harmonics or vibrations are those that use genetic [20], random or selective harmonic elimination algorithms (called calculated techniques) [21]-[28]. All of them provide good results but require a high degree of computation.

With a 3-level inverter structure, and when the inverter must operate with a low number of switches pulses per period, (i.e. high-power inverters for low switch losses), the THD value is high. To achieve a reduction in the THD value and reduce the acoustic vibration, techniques with harmonics injection and frequency modulated triangular carrier can be used [29]-[30].

The usefulness of decoupling M, from the electrical spectrum of the output wave is evidenced in this type of applications. While in a classic SPWM technique, the frequency of the carrier wave is constant and its value is M times $\omega_m$, the modulation presented by [29]-[31] allows the maximum instantaneous frequency of the carrier wave to be fixed in the interval [M, 2M] times $\omega_m$.

Efficient electrical and acoustic results were presented using a linear modulation law to change the carrier frequency, (HIPWM-FMTC2) [32]. This solution eliminates the restriction on the maximum value of the instantaneous frequency modulation order inherent in [31], increasing it up to the switching speed limit of the insulated-gate bipolar transistors (IGBTs). In this case, the maximum instantaneous carrier wave frequency will be in the range of [2M, L] where L is the theoretical limit of the switching speed of the power inverter. The procedure is based on cancelling the output signal switching when the slope of the modulating signal is small (close to the maximum and minimum values). In return, for elevated slopes of this one, the value of the instantaneous frequency of the carrier wave is increased. This change shift the most significate harmonics to higher frequencies. The problem with the law of linear frequency modulation is the complexity of formulating the output wave using the Fourier series method.

This problem disappears using a sinusoidal modulation of the carrier wave frequency. A truncate square-cosine has been chosen because it presents the best results. The main advantage of this modulation function is that it permits obtaining mathematically the output voltage using the Fourier series method. This feature is very useful for the design of electrical machines using simulators, which analyze the mechanical and thermal behavior as a function of the applied electrical voltage wave.

In addition, the instantaneous frequency is higher for longer. As a result, the side bands of the switching frequencies have a better distribution (fewer terms, with less amplitude and closer to the central frequencies), making it easier to locate the optimum control parameters for each machine and load level. This procedure will be called Harmonic Injection Pulse Width Modulation with Frequency Modulation Triangular Carrier 3 technique, HIPWM-FMTC3. The adjustment of the values of the control parameters modifies the electrical spectrum of the output wave for a given number of pulses per period, M, and offers the possibility to avoid certain terms of this spectrum, which can produce mechanical resonances, or which present a high winding factor for the implemented induction motor.

The paper will be divided into the following sections: section 2 of the document examines mechanical vibrations of electromagnetic origin in induction machines and their relationship with their mechanical design. The technique is presented in Section 3. Section 4 develops the mathematical formulation and section 5 provides experimental results with a description of the application to verify the usefulness of the technique.

## 1. Electromagnetic vibration on the induction motors

The use of alternating current motors in electric vehicles has given a strong impulse to the study of new (multi-levels) power inverters topologies to increase efficiency and reduce vibrations originating from the modulation technique used to control the inverter. Mechanical vibrations of electromagnetic origin appear at frequencies linked to the temporal harmonics produced by the inverter. Increasing M increases switching losses and decreases battery efficiency. Furthermore, this procedure does not guarantee that the mechanical resonances at high frequencies can be activated. To improve drive performance without reducing the quality of the voltage applied to the motor, it is useful to reduce the value of $\overline{M}$ and decouple it from the electrical spectrum. Radial magnetic vibration on the induction motors is caused by the axial deformation of the rotor and stator due to the products between the harmonic terms of the flow densities as derived from Maxwell stress tensor.

The amplitude of the radial magnetic pressure density [N/m$^2$] for each vibration order depends on the harmonics of the magnetic flow density in the air gap. This undesirable effect is produced by: 1) the interaction between the harmonic terms of the stator at a given frequency; 2) the interaction between the harmonic terms of the stator and the rotor at different frequencies; and, 3) the interaction between the rotor harmonic terms at the same frequency. These harmonics will be spatial (consequence of the machine design) and temporal (consequence of the modulation technique). The most important source for magnetic vibrations is the product of the harmonic terms of magnetic flux densities produced by the stator and rotor currents. Therefore, the absence of large-amplitude harmonic terms in the voltage supply allows the level of vibration produced by the motor to be reduced.

A modulation technique with low amplitude harmonic terms will reduce their RMS levels. Therefore, the use of a technique that allows harmonics spectra to be redistributed will avoid mechanical resonances and interactions between spatial and temporal harmonics. The reduction in the vibration level contributes to improve the reliability and service life of the motor.

## 2. Description of the proposed technique

As discussed in the introductory section, the key of the technique is based on modifying the frequency of the carrier wave in sync with the modulating wave. The number of commutation pulses to form the output wave is increased in the interval wherein the modulating signal has the greatest slope (taking as limit the maximum switching speed of the inverter). The commutation pulses are cancelled during the time intervals in which the modulating wave is closest to its maximum or minimum amplitude (when it has less slope). Thus, the instantaneous pulsation of the carrier signal $\omega_i$, will be a synchronized discontinuous function of the fundamental term of the modulating wave, which is defined as follows:

$$\omega_i = \frac{d\theta}{dt} = A_M \omega_m [\cos^2 \omega_m t - K] \quad (1)$$

In those intervals wherein (1) take negative values, $\omega_i$ will be set to zero. The modulating wave and $\omega_i$ must be synchronized functions as shown in Fig. 1. The parameter K (truncated level) is defined as a real number in the interval [0,1]. $A_M$ is a parameter for control of the maximum frequency of the carrier wave. These parameters make it possible to modify the electrical spectrum without changing M. The relative frequency of the carrier and modulating waves is the instantaneous frequency modulation order:

$$M(t) = \frac{\omega_i}{\omega_m} = A_M [\cos^2 \omega_m t - K] \quad (2)$$

The average value of $M(t)$ defines the number of commutation pulses per fundamental period, $T_m$:

$$\overline{M} = \frac{1}{T_m} \int_0^{T_m} A_M [\cos^2(\omega_m t) - K] dt \quad (3)$$

where $T_m = \frac{2\pi}{\omega_m}$. The first zero for equation (1) is reached for the time $t_1 = \frac{\cos^{-1}\sqrt{K}}{\omega_m}$, and thus:

$$\overline{M} = 4 \frac{\omega_m}{2\pi} \int_0^{t_1} A_M [\cos^2(\omega_m t) - K] dt =$$

$$= \frac{A_M}{\pi} \left\{ \frac{\sin(2\omega_m t_1)}{2} + (1 - 2K)\omega_m t_1 \right\} \quad (4)$$

$\overline{M}$ is the average value of the frequency modulation order, and it is proportional to the area under the curve in figure 1.b), according with (3).

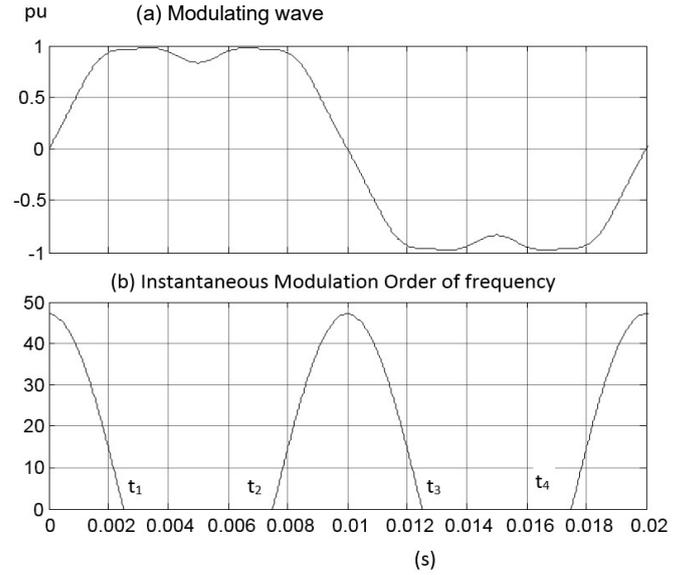

Fig.1. (a) Modulating wave versus time, (b) Instantaneous modulation order versus time ($f_m$=50 Hz, K= 0.5 and $A_M = 30\pi$, $\overline{M}$ =15).

To achieve the synchronization between the modulating and carrier waves, $\overline{M}$ must be set to an integer value. For avoid pair harmonics and triple harmonics with 3-phases inverters, $\overline{M}$ should be set to an odd value multiple of 3. For each odd value of $\overline{M}$ multiple of 3, infinite pairs of K and $A_M$ can be selected (2), which allows to define the synchronization between the modulating and the carrier waves. The purpose of this synchronization is that the maximum instantaneous value of the carrier frequency coincides with that of the instant in which the sinusoidal modulating wave has greater slopes, (angles close to 0 and π rad). It should then decline following a square cosine function, around those phases and reaches zero during a time interval that depends on K as it can be seen on (1) and (2).

For K = 0, the frequency of the carrier wave will theoretically be zero only for the phase values of the modulating wave: π/2 and 3π/2 (0.005 and 0.015 s for 50 Hz). The same carrier wave could be obtained by using [30], with $\omega_c = 30\omega_m$ and $k_f = 0$, being $\omega_c$ and $k_f$ the characteristic parameters for the technique referred. Therefore, the voltage obtained at the inverter output would be the same in both techniques and the excursion of the instantaneous pulsation will be $30\omega_m$. For any other K value below 1, the carrier frequency of the proposed technique will be zero not only at the angles π/2 and 3π/2 of the modulating wave, but also at two-time intervals around these phases, with the advantage of increasing the instantaneous frequency of the carrier wave around the phases 0 and π rad of the modulating wave. The modulating wave presents the maximum slope and changes faster for these angles; so, it is convenient to increase the number of switching pulses around them.

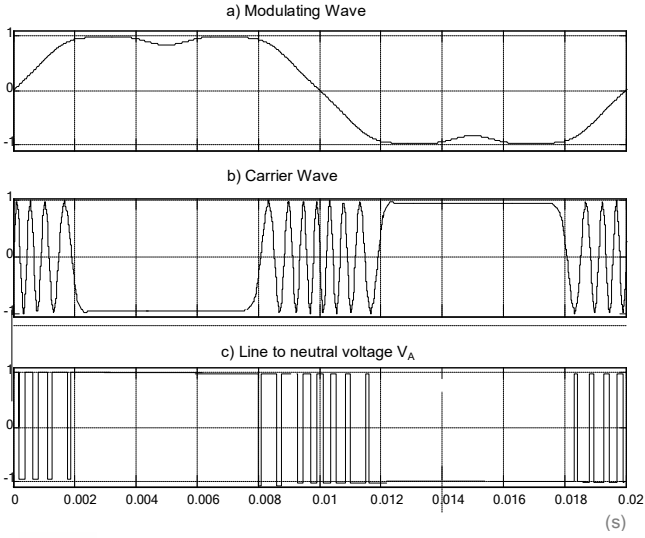

*Fig. 2. (a) Modulating wave for f=50 Hz; (b) Carrier signal HIPWM-FMTC3 technique (K = 0.5 and $A_M = 30\pi$) and (c) Line to Neutral Voltage at the output inverter.*

Fig. 2 shows the characteristic wave shapes for a pair of $A_M$ and K values required to obtain an $\overline{M}$ value of 15, calculated using Eq. (3) with K = 0.5 and $A_M = 30\pi$. For this K value, the inverter only switches during one half of the total period of the modulating wave of 20 ms (50 Hz). The maximum excursion of the carrier frequency will be $A_M \cdot \omega_m \cdot (1-K) = 30\pi\omega_m \cdot 0.5$ as is deduced from (1)-(4). Therefore, $t_1$ will be 2.5 ms; and $t_2$, $t_3$ and $t_4$ will be 7.5, 12.5 and 17.5 ms respectively. The pulses are concentrated around the instants mT/2, being T, the fundamental period of the modulating wave, and m, the natural number series. At those instants, the instantaneous frequency of the carrier wave is the maximum: $(30\pi/2)$ times the modulating wave frequency, instead of 15 as would be the case in a SPWM technique, and the area under the curve of the instantaneous modulation order in Fig. 1(b) is $\overline{M}$ = 15. Each value of K will allow a different value of $A_M$ when $\overline{M}$ is set.

From equations (3)-(4), Table I shows how some variables evolve versus K when $\omega_m$ and $\overline{M}$ are adjusted to $100\pi$ rad/s and 15, respectively. The variables presented are: $t_1$, which represents a quarter of the time in which commutations are allowed per period of the output wave (see Figure 1); $A_M$, for adjusting the maximum frequency of the carrier wave; and the central harmonic order, which is calculated as $A_M \cdot (1-K)$. The central frequency is $A_M \cdot (1-K) \cdot \omega_m$.

TABLE I
VALUES OF VARIABLES $A_M$, $t_1$ AND CENTRAL HARMONIC ORDER VERSUS K

| K | $A_M$ | $t_1$ (ms) | $A_M(1-K)$ |
|---|---|---|---|
| 0.2 | 44.277 | 3.524 | 35.422 |
| 0.3 | 55.134 | 3.155 | 38.594 |
| 0.4 | 70.638 | 2.820 | 42.383 |
| 0.5 | 30·π | 2.5 | 47.124 |
| 0.6 | 133.513 | 2.180 | 53.405 |
| 0.7 | 208.142 | 1.845 | 62.443 |
| 0.8 | 386.859 | 1.476 | 77.372 |

It is advisable not to increase K above a certain upper limit due to the maximum switching speed limitation of the inverter IGBTs. For the tests carried out, K has been limited to 0.7.

### 3. Mathematical expression of the output signal: results and discussions

The expression is developed in terms of the Fourier series of the line to neutral voltage $V_A(t)$ obtained with this new HIPWM-FMTC3 technique using Fig. 3:

$$\frac{a_0}{2} = \frac{1}{2\pi}\int_0^{2\pi} f(\omega_i t) d\omega_i t = \frac{1}{2\pi}\left[\int_{-\pi}^{-\alpha}\left[-\frac{E}{2}\right]d\omega_i t + \int_{-\alpha}^{\alpha}\left[\frac{E}{2}\right]d\omega_i t + \int_{\alpha}^{\pi}\left[-\frac{E}{2}\right]d\omega_i t\right] = \frac{E}{2}\left(2\frac{\alpha}{\pi}-1\right)$$

(5)

being E, the DC-LINK level.

$$a_n = \frac{1}{\pi}\int_0^{2\pi} f(\omega_i t)\cos(n\omega_i t)d\omega_i t =$$
$$\frac{1}{\pi}\left[-\int_{-\pi}^{-\alpha}\left[\frac{E}{2}\right]\cos(n\omega_i t)d\omega_i t + \int_{-\alpha}^{\alpha}\left[\frac{E}{2}\right]\cos(n\omega_i t)d\omega_i t - \int_{\alpha}^{\pi}\left[\frac{E}{2}\right]\cos(n\omega_i t)d\omega_i t\right] = \frac{4E}{2\pi n}\sin(n\alpha)$$

(6)

$$b_n = \frac{1}{\pi}\int_0^{2\pi} f(\omega_i t)\sin(n\omega_i t)d\omega_i t =$$
$$\frac{1}{\pi}\left[-\int_{-\pi}^{-\alpha}\left[\frac{E}{2}\right]\sin(n\omega_i t)d\omega_i t + \int_{-\alpha}^{\alpha}\left[\frac{E}{2}\right]\sin(n\omega_i t)d\omega_i t - \int_{\alpha}^{\pi}\left[\frac{E}{2}\right]\sin(n\omega_i t)d\omega_i t\right] = 0$$

(7)

The value of α is obtained from the intersection of the modulating and carrier functions (see fig. 3):

$$H(\omega_m t) = 1.15\cos(\omega_m t) - 0.27\cos(3\omega_m t) - 0.029\cos(9\omega_m t) = \left(2\frac{\alpha}{\pi}-1\right) \to \alpha = \frac{\pi}{2} + \frac{\pi}{2}H(\omega_m t)$$

(8)

being $H(\omega_m t)$, the modulating function. By replacing the expression for α in the Fourier decomposition for a periodic signal, the line to neutral voltage at output $V_A(t)$ can be obtained as:

$$V_A(t) = \frac{a_0}{2} + \sum_{n=1}^{\infty} a_n \cos(\omega_i t) + \sum_{n=1}^{\infty} b_n \sin(\omega_i t) =$$
$$= \left(\frac{E}{2}\right) H(\omega_m t) +$$
$$+ \frac{4}{\pi}\left(\frac{E}{2}\right)\sum_{n=1}^{\infty}\frac{1}{n}\sin\left(\frac{\pi}{2}+\frac{\pi}{2}H(\omega_m t)\omega_m t\right)\cos(n\omega_i t) \quad (9)$$

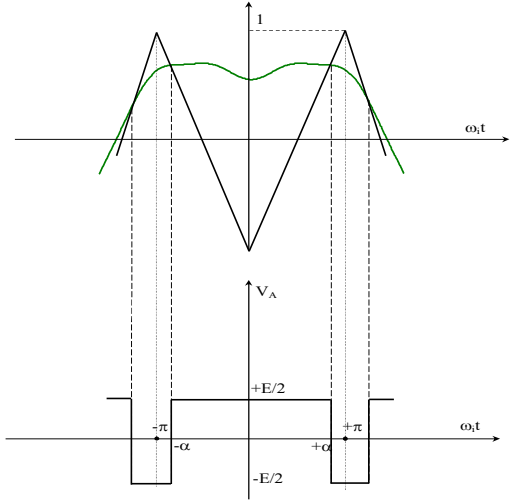

Fig. 3. HIPWM-FMCT3 modulation.

The first summand represents the fundamental term while the second summand represents the harmonic terms of the output wave. The instantaneous phase of (1) is:

$$\omega_i t = \theta = \int [A_M(\cos^2(\omega_m t) - K)]\,dt = \int \frac{A_M}{2}[1+$$
$$+\cos(2\omega_m t) - 2K]\,dt = A_M\left[(0.5-K)t + \frac{\sin(2\omega_m t)}{4\omega_m}\right]$$
(10)

By substituting (10) in (9), for the time interval were M(t) is not null, this is: $t_2 > t > t_1$ and $t_4 > t > t_3$, (see fig 6 and (1)), the line-to-neutral voltage $V_A(t)$ is defined as:

$$V_A(t) = \left(\frac{E}{2}\right)H(\omega_m t) +$$
$$+ \frac{4}{\pi}\left(\frac{E}{2}\right)\sum_{n=1}^{\infty}\frac{1}{n}\sin\left(\frac{n\pi}{2}\left(\frac{n\pi}{2}+\frac{n\pi}{2}H(\omega_m t)\right)\right) \cdot$$
$$\cdot \cos n\left(A_M(K-0.5))t + \frac{\sin(2\omega_m t)}{4\omega_m}\right) \quad (11)$$

$$t_2 = \frac{(\frac{\pi}{2}+\cos^{-1}\sqrt{K})}{\omega_m} > t > t_1 = \frac{(\frac{\pi}{2}-\cos^{-1}\sqrt{K})}{\omega_m}$$
$$t_4 = \frac{(\frac{3\pi}{2}+\cos^{-1}\sqrt{K})}{\omega_m} > t > t_3 = \frac{(\frac{3\pi}{2}-\cos^{-1}\sqrt{K})}{\omega_m}$$

For the interval were M(t) is null, $V_A(t)$ will be:

$$V_A(t) = 1 \to t \leq t_1$$
$$V_A(t) = -1 \to t_2 \leq t \leq t_3 \quad (12)$$
$$V_A(t) = 1 \to t \geq t_4$$

Therefore, the widths of the side bands in the harmonic spectrum of the output voltage will also increase (i.e., a greater number of harmonic terms, but of lower amplitudes), as n increases according to Fourier series (Fig. 4), where harmonic spectra from K = 0.3 to 0.8 have been highlighted.

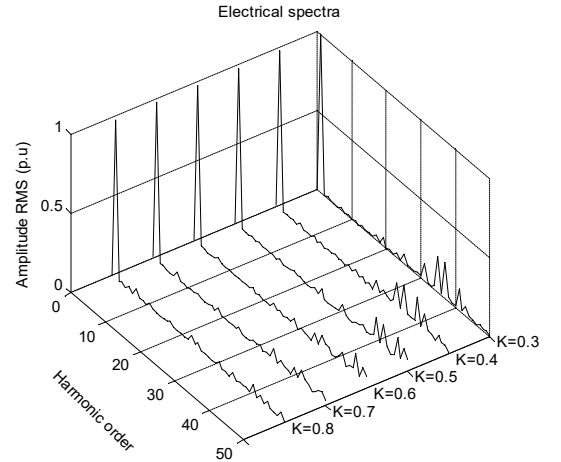

Fig. 4. Frequency spectra line-to-line voltage $V_{AB}$ ($\overline{M}$=15, K = 0.3 to K=0.8)

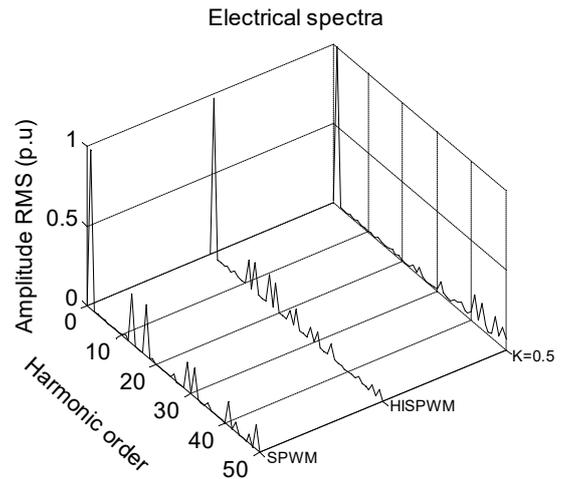

Fig. 5. Spectra comparison of the line-to-line voltage $V_{AB}$ (HIPWM-FMTC3 technique, $\overline{M}$=15, K = 0.5); Harmonic Injection SPWM and SPWM techniques.

Fig. 5 shows the improvement in the frequency spectrum of voltage achieved by the proposed technique compared to two classic techniques (SPWM and Harmonics Injection SPWM) for M=15. It may be noted that, for K=0.5, the first relevant term (LOH) is located at a frequency 23 times higher than the switching frequency.

By using equations (11)-(12), voltage $V_A(t)$, i.e. line to neutral voltage, is depicted in Fig. 6, while the corresponding frequency spectrum is included in Fig. 7; in both figures, the values of the parameters are: K = 0.5, and $A_M = 30\pi$. Because the maximum shift of the carrier frequency will be $A_M \cdot \omega_m \cdot (1-K)$ according to (3), $15\pi\omega_m$ is the new central frequency for the first order side bands.

For a classical SPWM technique and for $\overline{M}$ =15, the first side bands are found around the 15th harmonic, as may be seen in Fig. 8.

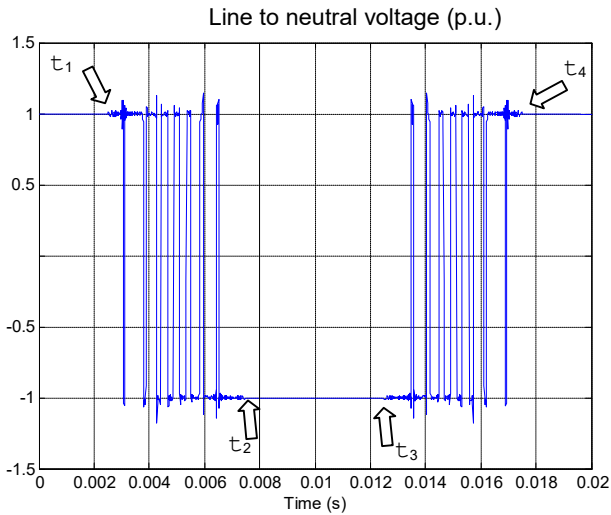

Fig. 6. $V_A(t)$ obtained using the proposed technique, (K = 0.5, $A_M = 30\pi$, $\overline{M}$ = 15, $\omega_m = 100\pi$ rd).

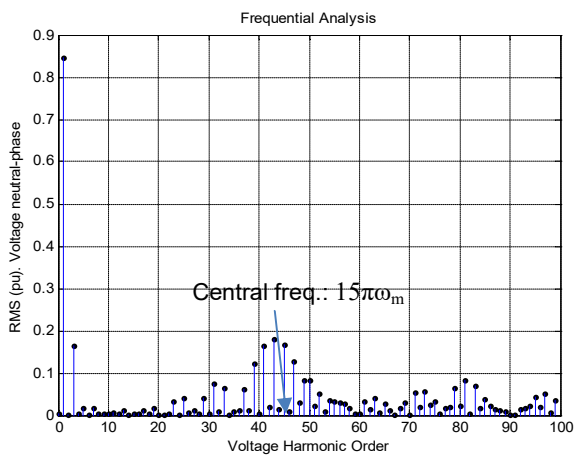

Fig. 7. Frequency analysis of the proposed technique: pu voltage line-to-neutral $V_A(t)$ ($\overline{M}$ =15, K=0.5, $A_M=30\pi$) and central frequency: $0.5 \cdot 30\pi \ \omega_m$

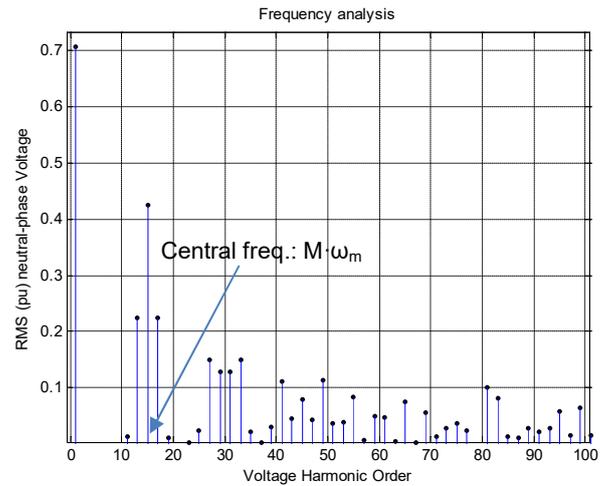

Fig. 8. Frequency analysis of the SPWM technique: pu voltage line-to-neutral $V_A(t)$ (M=15).

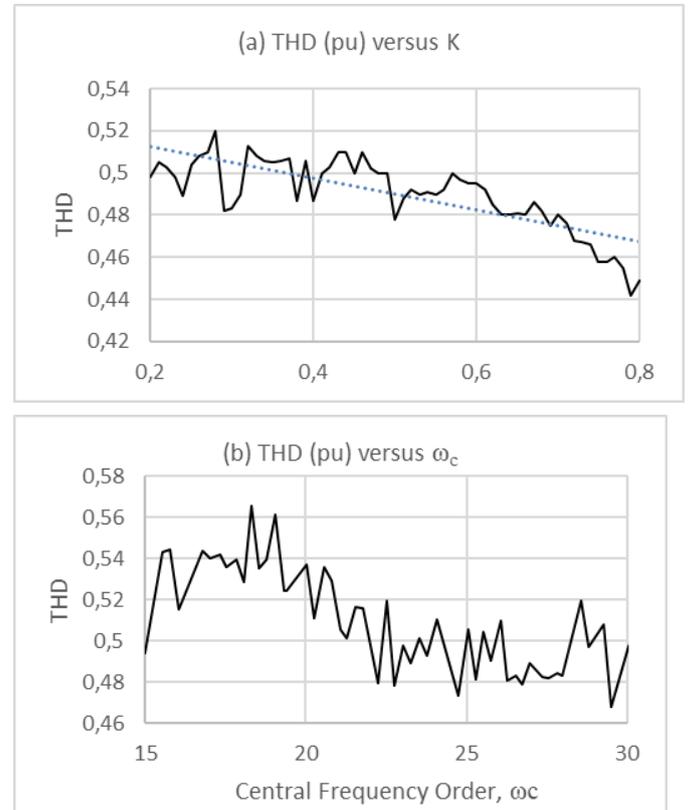

Fig. 9. THD values (pu) (a) versus K for HIPWM-FMTC3 technique; (b) versus control parameter $\omega_c$, for HIPWM-FMTC technique.

Figures 9(a) and 10(a) display the behavior of THD and DF respectively for the HIPWM-FMTC3 technique versus the control parameter K. On the other hand, Figures 9.b and 10.b show the same factors for the HIPWM-FMTC [31] technique in relation to its control parameter, $\omega_c$ which can vary between M and 2M. In both figures a relative reduction of both factors can be observed for certain values of parameter K in comparison with [31]. The analytical expressions used to calculate THD and DF are included in [33].

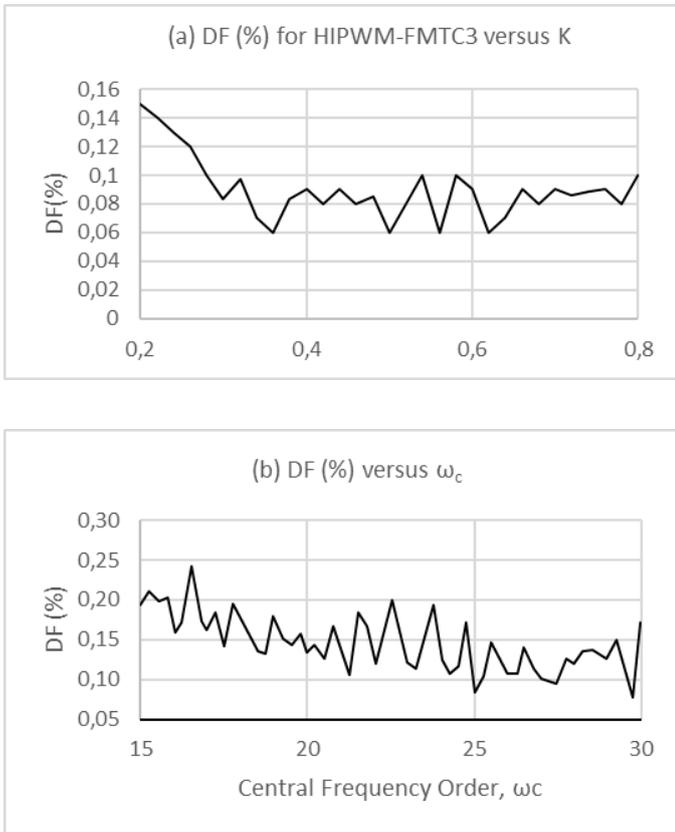

*Fig. 10. Distortion Factor, DF (%), (a) versus K for HIPWM-FMTC3 technique; (b) versus control parameter $\omega_c$, for HIPWM-FMTC technique.*

TABLE II
FUNDAMENTAL TERM LINE-TO-LINE, AND THD VERSUS K (SIMULATION)

| Modulation strategy | $V_1$ (pu) | Voltage THD (%) |
|---|---|---|
| HIPWM-FMTC3 (K = 0.7) | 0.66 | 54.83 |
| HIPWM-FMTC3 (K = 0.6) | 0.70 | 57.81 |
| HIPWM-FMTC3 (K = 0.5) | 0.84 | 54.03 |
| HIPWM-FMTC3 (K = 0.3) | 0.72 | 52.94 |
| HIPWM-FMTC3 (K = 0.2) | 0.73 | 40.75 |
| HIPWM-FMTC $\omega_C = 24.75\omega_m$; $k_f = 19.5\omega_m$ | 0.75 | 47.30 |
| SPWM (M=15) | 0.65 | 70.23 |

Table II compares the simulation results (relative to the DC converter voltage p.u.) for the fundamental term of the line to line voltage and the THD using: the HIPWM-FMTC3 technique for different K values; the HIPWM-FMTC technique [29-31], adjusted for the higher fundamental term of the output wave, and the SPWM technique.

## 4. Experimental Results

To validate the usefulness of the proposed technique, it has been applied to control a 3-levels, 3-phases Skiip 132GDL 120-412 CTV open-loop VSI inverter by Semikron that it drives an induction motor 230/400 V line-to-line, with a rated power of 600 W, 36 stator slots, 30 rotor bars and 2 poles pair. The line-to-neutral voltage at the inverter output was 234 V RMS (fundamental term) for all measurements made with the different values of parameter K. This voltage value was the reference value for direct supply from the grid. All measurements were carried out with the induction motor without load.

The experimental configuration is shown in Fig. 11. It consists of a PC-DSP that generates control signals over the VSI. The output voltage was measured with a TeamWare Equa network analyzer, which provides a spectrum with the first 50 voltage harmonics, as well as the first 50 current harmonics. Vibration transmitted on the housing was measured and analyzed with a 01-dB METRAVIT Symphonie model sonometer (m/s$^2$ RMS).

Fig. 12 (a) shows the line to neutral voltage, $V_A$ (t) when the motor is fed from the power inverter in the experimental setup, with K=0.5, $A_M$ = 30π and $\overline{M}$ = 15, while Fig. 12 (b) illustrates its electrical spectrum. The harmonic terms around the first sideband can be observed.

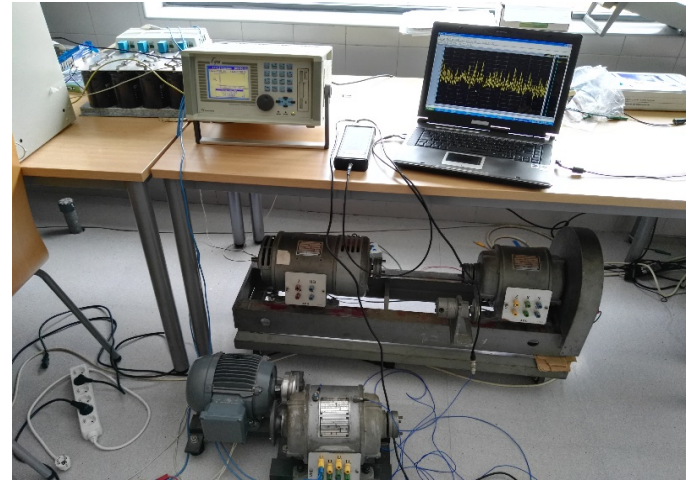

*Fig. 11. Induction motor used and laboratory assembly for vibration measurements.*

Electromagnetic vibrations in the motor housing are caused by the generation of magnetomotive forces (FMM). The RMS value of magnetic flux density and its decomposition in harmonic terms depends on the frequencies and amplitudes of these FMM. Likewise, the electromagnetic radial forces produced due to them will depend on the harmonic decomposition of the magnetic flux density. When a motor is connected to the grid (without power inverter), the amplitudes and frequencies of the vibration waves transmitted to the housing depend primarily on the winding distribution factor for each harmonic of the stator slot, and the harmonics of the slot, and these depend on the number of rotor bars.

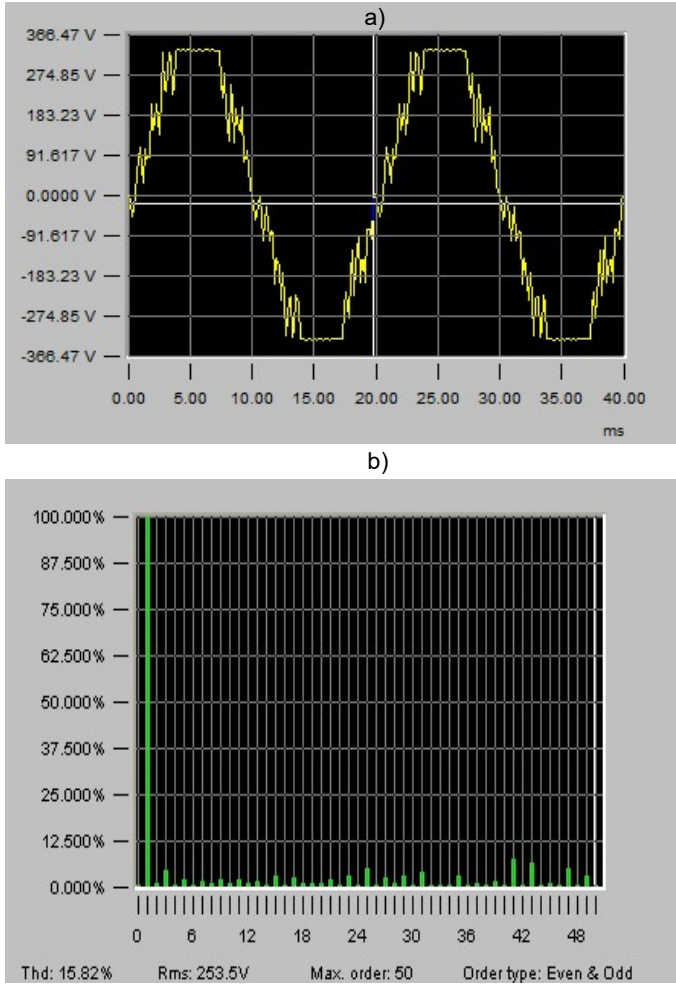

Fig. 12. (a) Line-neutral voltage $V_A(t)$, using HIPWM-FMTC3 technique (K =0.5, $A_M = 30\pi$) measured with the Electric Power Quality Analyzer Equa; (b) Frequency spectrum.

Fig. 13.a) allows the identification of these space harmonics (slot harmonics) with central frequencies: $f_r$=n·f·(rotor bars)/(poles pair), that is, 15nf·(1-s), (15f for n=1, 30f for n=2); lateral band frequencies: 15n·(1-s)±2 (13f and 17f for n=1) being $f_r$ the frequency of the vibrations, n=0,1,2..., s the slip, (which can be approximated to 1), and f=50 Hz. Another group of spatial harmonics that can be expected are those produced by products between the terms of magnetomotive forces of the rotor, with frequencies: $f_r$=2·(1±15n); (28 and 32f for n=1, and 58f and 62f if n=2) [34]. These terms appear for 650, 750, 850, 1400, 1500, 1600 Hz (these are the harmonics: 13, 15, 17, 28, 30 and 32). With the technique presented the activation of these spatial harmonics can be avoided for $\overline{M}$=15.

Fig. 13.b) show the mechanical vibrations produced by the interaction of the stator and rotor harmonics when the motor is supplied with an inverter controlled by the proposed technique. Likewise, Fig. 13.c) illustrates the mechanical vibrations measured by the accelerometer for the technique SPWM technique. The frequency terms, $f_r$ = f ± f·(15n-(1-s) ± 1) [34] are appreciable. These are activated with SPWM strategy (Fig.13.c) but with the proposed technique its activation can be avoided (Fig.13.b).

The measured vibration level with HIPWM-FMTC3 is about 4 dB (referred to 1e-6 m/s$^2$) lower than the measured level with a PWM classical technique. The difference between both techniques implies a decrease of up to 4.5 dB in the vibration level with the appropriate value of K using this motor and $\overline{M}$=15.

The lowest level of vibrations is achieved with a value of K=0.465, but this value depends on each motor, by the interaction of spatial and temporal harmonics.

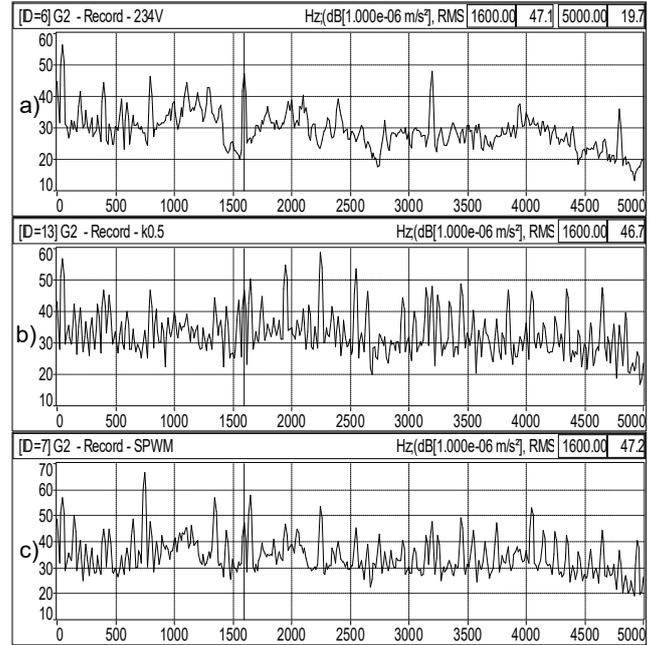

Fig. 13 (a) Vibration frequency spectrum for sinusoidal power supply, (b) Vibration frequency spectrum for HIPWM-FMTC3 (K=0.5, $A_M = 30\pi$), (c) Vibration frequency spectrum for SPWM.

Table III compares the vibrational levels measured with a sinusoidal feed, and with the inverter controlled by the HIPWM-FMTC3 technique ($\overline{M}$=15) for different K values and last but not least with the SPWM technique.

TABLE III
VIBRATION LEVEL VERSUS K (MEASURED VALUES)

| Modulation strategy | RMS (dB) |
|---|---|
| Sinusoidal supply | 60.0 |
| HIPWM-FMTC3 (K = 0.7) | 67.2 |
| HIPWM-FMTC3 (K = 0.6) | 65.1 |
| HIPWM-FMTC3 (K = 0.5) | 64.5 |
| HIPWM-FMTC3 (K = 0.3) | 66.0 |
| HIPWM-FMTC3 (K = 0.2) | 65.9 |
| SPWM (M=15) | 68.8 |

## 5. Conclusion

A new generated type PWM technique for a 3-level inverter was presented. The mathematical expression for the line-to-neutral voltage $V_A$ of the inverter output was described. An increase in the fundamental term of the output voltage can be observed, along with a reduction of the harmonic terms amplitude. Due to the particularity of using a truncated cosine law to adjust the carrier wave frequency, the instantaneous frequency of the output wave may increase significantly compared to a classical technique for a given number of pulses. As a result, harmonics clusters are shifted to higher frequencies.

The change in the values of the control parameters K and $A_M$ (for a given $\overline{M}$ value) shifts the frequency spectrum of the output voltage. This feature could be very useful when feeding an induction motor to avoid resonance frequencies, or to avoid stator or rotor MMF harmonics with high winding factor.

In addition, by using a truncated cosine variation law of the carrier wave frequency instead of a cosine one (HIPWM-FMTC), or a linear one (HIPWM-FMTC2), the instantaneous frequency is higher for longer, allowing a more favorable evolution of the electrical spectrum. Also, the THD factor is reduced with respect to other techniques analyzed: SPWM, HIPWM-FMTC and finally, HIPWM-FMTC2. The sideband terms of the electrical frequency spectrum were also divided and dispersed, reducing the amplitude of each term to attenuate the vibration level of an induction motor.

It has been justified that the electric spectrum can be decoupled from the order of frequency modulation $\overline{M}$, and that the optimum result can be imposed in terms of vibration level, fundamental term amplitude, or the THD level for any imposed $\overline{M}$ value.

## 6. Acknowledgments

This manuscript has been funded by: SPANISH NATIONAL RESEARCH PLAN (2015-2017) with the number: ENE2013-46205-C5-5-R.

## 7. References

[1] J. I. Leon, S. Kouro, L. G. Franquelo, J. Rodriguez, and B. Wu,"The essential role and the continuous evolution of modulation techniques for voltage-source inverters in the past, present, and future power electronics," IEEE Trans. Ind. Electron, 2016, 63, (5), pp. 2688–2701

[2] Z. Zhang, O.C. Thomsen, and M.A. Andersen, "Discontinuous PWM modulation strategy with circuit – level decoupling concept of three-level neutral-point-clamped (NPC) inverter," IEEE Trans. Ind. Electron., 2013, 60, (5), pp. 1897-1906

[3] G. Grandi and J. Loncarski, "Simplified implementation of optimized carrier-based PWM in three-level inverters," IET Electronics Letters, 2014, 50, (8), pp. 631–633

[4] A.C. Binoj Kumar and G. Narayanan, "Variable-Switching frequency PWM technique for Induction motor drive to spread acoustic noise spectrum with reducen current ripple", IEEE Transactions on Industry Application, 2016, 52, (5), pp. 3927-3937

[5] G. Narayanan, D. Zhao, H. K. Krishnamurthy, R. Ayyanar and V. T. Ranganathan, "Space vector based hybrid PWM techniques for reduced current ripple," IEEE Transactions on Industrial Electronics, 2008, 55, (4), pp. 1614–1627

[6] A. Choudhury, P. Pillay and S.S. Williamson, "A hybrid PWM-based DC-link voltage balancing algorithm for a three-level NPC DC/AC traction inverter drive," IEEE Journal of Emerging and Selected Topics in Power Electronics, 2015, 3, (3), pp. 805–816

[7] J. Holtz and N. Oikonomou, "Optimal control of a dual three-level inverter system for medium-voltage drivers," IEEE Trans. Ind. Appl., 2010, 46, (3), pp. 1034-1041

[8] M. Dahidah, G. Konstantinou, and V. Angelidis, "A review of multilevel selective harmonic elimination PWM: Formulations, solving algorithms, implementation and applications," IEEE Trans. Power Electron., 2015, 30, (8), pp. 4091-4106

[9] J. Rodriguez et al., "Multilevel converters: An enabling technology for high-power applications," Proc. IEEE, 2009, 97, (11), pp. 1786-1817

[10] S. Kouro et al., "Recent advances and industrial applications of multilevel converters," IEEE Trans. Ind. Electron., 2010, 57, (8), pp. 2553-2580

[11] R. Ramchand, K. Sivakumar, A. Das, C. Patel and K. Gopakumar, "Improved switching frequency variation control of hysteresis controlled voltage source inverter-fed IM drives using current error space vector," *IET Power Electronics*, vol. 3, no. 2, , Mar. 2010: 219–231.

[12] M.S.A. Dahidah, G.S. Konstantinou and V.G. Agelidis, "Selective harmonic elimination pulse-width modulation seven-level cascaded H-bridge converter with optimized DC voltage levels," IET Power Electronics, 2012, 5, (6), pp. 852–862

[13] S. Bifaretti, L. Tarisciotti, A. Watson, P. Zanchetta, A. Bellini and J. Clare, "Distributed commutations pulse-width modulation technique for high-power AC/DC multi-level converters," IET Power Electronics, 2012, 5, (6), pp. 909–919

[14] P. Palanivel and S. S. Dash, "Analysis of THD and output voltage performance for cascaded multilevel inverter using carrier pulse width modulation techniques," IET Power Electronics, 2011, 4, (8), pp. 951–958

[15] C. Tan, D. Xiao, J. E. Fletcher and M. F. Rahman, "Carrier-based PWM methods with common-mode voltage reduction for five-phase coupled inductor inverter," IEEE Transactions on Industrial Electronics, 2016, 63, (1), pp. 526–537

[16] H. Stemmler and T. Eilinger, "Spectral analysis of the sinusoidal PWM with variable switching frequency for noise reduction in inverter-fed induction motors," Power Electronics Specialist Conference, Taipei, Taiwan, June 1994, pp. 269–277.

[17] H. B. Ertan and N. B. Simsir, "Comparison of PWM and PFM induction drives regarding audible noise and

vibration for household applications," IEEE Transactions on Industry Applications, 2004, 40, (6), pp. 1621–1628

[18] F. Vargas-Merino, M. J. Meco-Gutierrez, J.R. Heredia-Larrubia, A. Ruiz-Gonzalez, "Low switching PWM strategy using a carrier wave regulated by the slope of a trapezoidal modulator wave," IEEE Transactions on Industrial Electronics, 2009, 56, (6), pp. 2270–2274

[19] F. Vargas-Merino, M. J. Meco-Gutierrez, J. R. Heredia-Larrubia and A. Ruiz-Gonzalez, "Highly efficient PWM strategy over FPGA," Electronic Letters, IET, 2008, 44, (24), pp. 1396–1398

[20] Quang-Tho Tran, Anh Viet Truong, Phuong Minh Le, "Reduction of harmonics in grid-connected inverters using variable switching frequency", International Journal of Electrical Power & Energy Systems, 2016, 82, pp. 242-251

[21] J. Napoles, J.I. Leon, R. Portillo L.G. Franquelo, and M.A. Aguirre, "Selective harmonics mitigation technique for high-power converters," IEEE Trans. Ind. Electron., 2010, 57, (7), pp. 2315-2323

[22] S.R. Pulikanti, M.S.A. Dahidah, and V. G. Agelidis, "Voltage balancing control of three-level active NPC converter using SHE-PWM," IEEE Trans. Power Del., 2011, 26, (1), pp. 258-267

[23] G. Konstantinou, M. Ciobotaru, and V.G. Agelidis, "Selective harmonics elimination pulse-width modulation of modular multilevel converters," IET Power Electron., 2013, 6, (1), pp. 96-107

[24] A. M. Trzynadlowski, F. Blaabjerg, J. K. Pedersen, R. L. Kirlin and S. Legowski, "Random pulse width modulation techniques for converter-fed drive systems- a review," IEEE Transactions on Industry Applications, 1994, 30, (5), pp. 1166–1175

[25] K.-S. Kim, Y.-G. Jung, Y.-C. Lim, "A new hybrid random PWM scheme," IEEE Transactions on Power Electronics, 2009, 24, (1), pp.192–200

[26] A. M. Trzynadlowski, K. Borisov and. L. Qin, "A novel random PWM technique with low computational overhead and constant sampling frequency for high-volume, low-cost applications," IEEE Transactions on Power Electronics, 2005, 20, (1), pp. 116–122

[27] J. Ruiz-Perez, A. Ruiz-Gonzalez, F. Perez-Hidalgo, "A new RPWM technique with harmonics injection and frequency randomized modulation," XIX International Conference on Electrical Machines, ICEM 2010, Rome, Italy, Sept. 1–6

[28] A. Guellal, C. Cherif, C. Larbes, D. Bendib, L. Hassaine, A. Malek, "FPGA based on-line Artificial Neural Network Selective Harmonic Elimination PWM technique", International Journal of Electrical Power & Energy Systems, 2015, 68, (6), pp. 33-43

[29] A. Ruiz-Gonzalez, M. Meco-Gutierrez; F. Perez-Hidalgo; F. Vargas-Merino; J. R. Heredia-Larrubia, "Reducing acoustic noise radiated by inverter-fed motors controlled by a new PWM strategy", IEEE Transactions on Industrial Electronics, 2010, 57, (1), pp. 228–236

[30] M. J. Meco-Gutierrez, F. Perez-Hidalgo, F. Vargas-Merino, J. R. Heredia-Larrubia, "A new PWM technique frequency regulated carrier for induction motors supply," IEEE Transactions on Industrial Electronics, 2006, 53, (5), pp. 1750–1754

[31] M. J. Meco-Gutiérrez; J. R. Heredia-Larrubia; F. Pérez-Hidalgo; A. Ruiz-González; F. Vargas-Merino; "Pulse width modulation technique parameter selection with harmonic injection and frequency modulated triangular carrier" IET Power Electronics, 2013, 6, (5), pp. 954-962

[32] A. Ruiz-Gonzalez, M. J. Meco-Gutierrez, F. Vargas-Merino, F. Perez-Hidalgo, J. R. Heredia-Larrubia, "Shaping the HIPWM-FMTC strategy to reduce acoustic noise radiated by inverter-fed induction motors," in XIX International Conference on Electrical Machines, ICEM 2010, Vilamoura, Portugal, Sept. 1–6

[33] M H Rashid, in Pearson "DC–AC Converters" in Pearson Ed: "Power Electronics", (Ed. Prentice Hall, 1988 first edn.), 4$^{th}$ edn. 2013, pp. 309

[34] J. F. Gieras, C. Wang, and J. Cho Lai, Inverter-Fed Motor, in Taylor & Francis "Noise of Polyphase Electric Motors" (CRC Press, 2006, 1st edn.), pp. 65-76